\documentclass[a4paper]{jpconf}
\usepackage{graphicx}
\usepackage{xspace}
\usepackage[a4paper,pdfborder={0 0 0}]{hyperref}
\usepackage[format=hang,labelfont=bf,hypcap=true]{caption}

\hypersetup{
  pdfauthor={Marek Schoenherr},
  pdftitle={Vector boson plus multijet production}
}
\newcommand{\aMCatNLO}{aM\protect\scalebox{0.8}{C}@N\protect\scalebox{0.8}{LO}\xspace}
\newcommand{\MCatNLO}{M\protect\scalebox{0.8}{C}@N\protect\scalebox{0.8}{LO}\xspace}

\newcommand{\POWHEG}{P\protect\scalebox{0.8}{OWHEG}\xspace}

\newcommand{\MEPS}{M\scalebox{0.8}{E}P\scalebox{0.8}{S}\xspace}
\newcommand{\NLOPS}{N\scalebox{0.8}{LO}P\scalebox{0.8}{S}\xspace}
\newcommand{\MENLOPS}{ME\protect\scalebox{0.8}{NLO}PS\xspace}
\newcommand{\MEPSatNLO}{M\scalebox{0.8}{E}P\scalebox{0.8}{S}@N\protect\scalebox{0.8}{LO}\xspace}
\newcommand{\BlackHat}{B\protect\scalebox{0.8}{LACK}H\protect\scalebox{0.8}{AT}\xspace}
\newcommand{\Sherpa}{S\protect\scalebox{0.8}{HERPA}\xspace}

\newcommand{\LO}{LO\xspace}
\newcommand{\NLO}{NLO\xspace}

\newcommand{\ATLAS}{ATLAS\xspace}
\newcommand{\CMS}{CMS\xspace}

\begin{document}
\title{\vspace{-3.5cm}{\normalsize IPPP/13/07, DCPT/13/14, LPN13-019}\\\vspace{2cm}
       Vector boson plus multijet production\vspace{5mm}}

\author{Marek Sch\"onherr}

\address{Institute for Particle Physics Phenomenology,
	 Durham University, Durham DH1 3LE, UK}

\ead{marek.schoenherr@durham.ac.uk}

\begin{abstract}
  In this contribution the developments in the description of vector 
  boson plus jets signatures at hadron colliders in recent years are 
  summarised. Particular focus is put on its relevance as background 
  to top physics.
\end{abstract}

\section{Introduction}

Within the first years of running of the LHC at centre-of-mass energies 
of 7 and 8 TeV top physics has been at the core of its physics programme. It 
serves both as a signal to be measured as precisely as possible and as 
a background to new physics, Higgs and Standard Model processes 
with many jets.

When considering measuring top quark production the production of $W$ and 
$Z$ bosons in association with many jets present the major backgrounds. 
In particular, the process $pp\to W+\ge 4$ jets constitutes an irreducible 
background to the semileptonic $t\bar t$ production channel. It therefore 
needs to be known with as large a precision as possible. Many advances 
have been made in recent years for this class of processes in their own 
right, reaching now a stage where they start to decrease the theoretical 
uncertainties for the relevant background channels below the leading order. 

While higher order calculations for $V(=W,Z)$ plus multiple jets become accessible 
they improve the description of QCD at large scales, leading to a 
stabilisation of the respective cross sections at the same time. But only 
when they are matched to parton showers a simultaneous and observable 
independent, reliable description of QCD dynamics at low scales is 
achieved. Such a matching offers additional benefits as the connection of 
low scale perturbative dynamics of the parton shower to the non-perturbative 
dynamics of hadronisation models and their subsequent hadron decays can be 
used to arrive at particle level descriptions that are directly comparable 
to experimental data. Multijet merging techniques can then be evoked to 
arrive at inclusive descriptions, combining successive multiplicities of 
fixed-order matrix elements matched to parton showers with their respective 
accuracies preserved and, at the same time, resumming multiscale logarithm 
associated with hierarchical multijet production.

\section{\protect\LO calculations and \protect\MEPS merging}

Multijet merging techniques at leading order accuracy are known for more 
than ten years \cite{Catani:2001cc,Mangano:2001xp,Lonnblad:2001iq,
  Krauss:2002up,Mangano:2006rw,Alwall:2007fs,Hoeche:2009rj,
  Hamilton:2009ne,Lonnblad:2011xx}. Combining tree-level matrix elements 
of successive parton multiplicities with parton showers into an inclusive 
description, preserving the respective accuracies, they are by now the 
work horses of the LHC experiments for multijet topologies. Their 
ability to describe data has been tested in various analyses 
\cite{Aad:2010ab,Aad:2012en,Aad:2011qv,Chatrchyan:2011ne,:2013is}.

\begin{figure}[t]
  \includegraphics[width=0.48\textwidth]{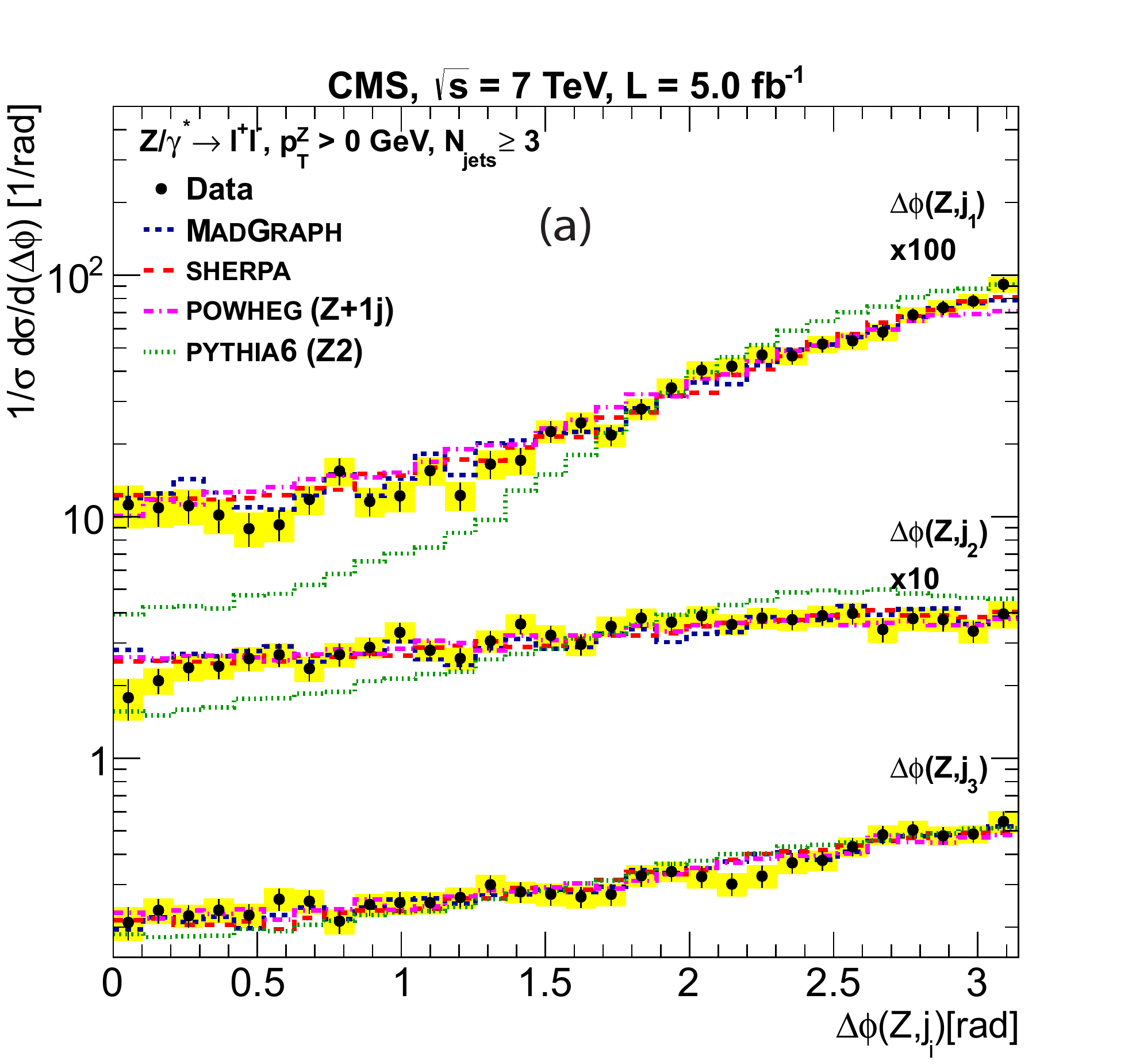}\hfill
  \includegraphics[width=0.48\textwidth]{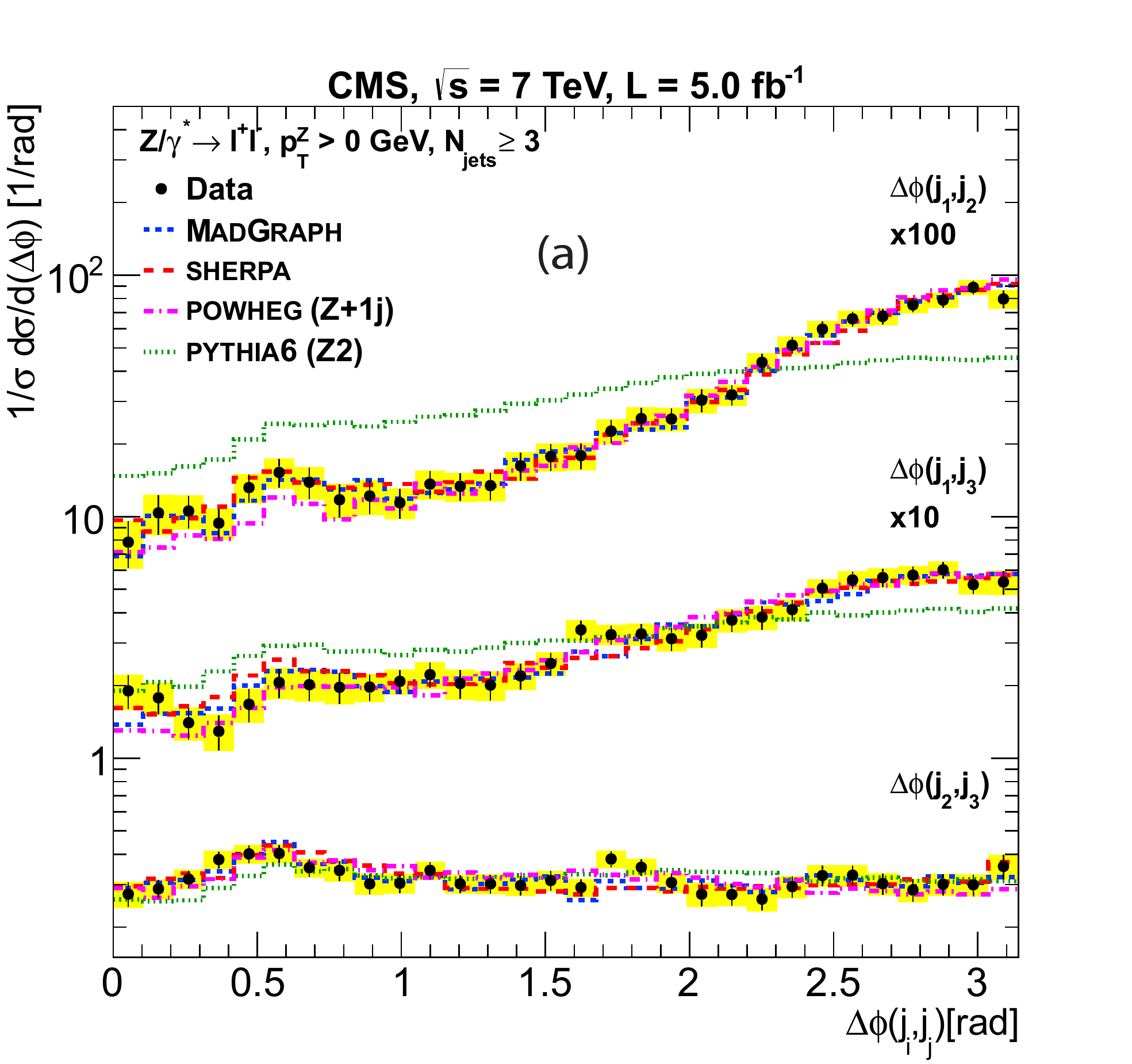}
  \caption{
	    Azimuthal decorrellation of the $Z$ and the 1st, 2nd and 3rd 
	    jet {\bf (left)} and of the 1st and 2nd, 1st and 3rd, and 2nd 
	    and 3rd jet in $pp\to Z+\ge 3$ jets at the LHC as measured by 
	    \CMS. Figures are taken from \cite{:2013is}.
	    \label{Fig:MEPS_CMS}
	  }
\end{figure}

Fig.\ \ref{Fig:MEPS_CMS} presents the most recent of these analyses wherein 
the ability of the \MEPS methods to describe the radiation pattern of the 
jets in $pp\to Z+\ge 3$ jets events is examined and good agreement is found. 
Nonetheless, the theoretical uncertainties of these methods are large owing 
to the leading order accuracy of the description of hard and/or wide angle 
parton emission, including quantum interference effects, only. By 
elevating their description to next-to-leading order the theoretical 
uncertainties on the respective observables can be reduced.

\section{\protect\NLO calculations, \protect\NLOPS and \protect\MEPSatNLO}

\subsection*{\protect\NLO}

While next-to-leading order calculations for $W$ and $Z$ production with 
up to two jets are known for some time \cite{Campbell:2002tg}, such 
calculation for $pp\to W+3$ jets \cite{Ellis:2009zw,Berger:2009zg}
and $pp\to W+4$ jets \cite{Berger:2010zx}, as are relevant as top backgrounds, 
have only become available recently. Similarly, $pp\to Z+3$ jets 
\cite{Berger:2010vm} and $pp\to Z+4$ jets \cite{Ita:2011wn} are available.

\begin{figure}[t]
  \includegraphics[width=\textwidth]{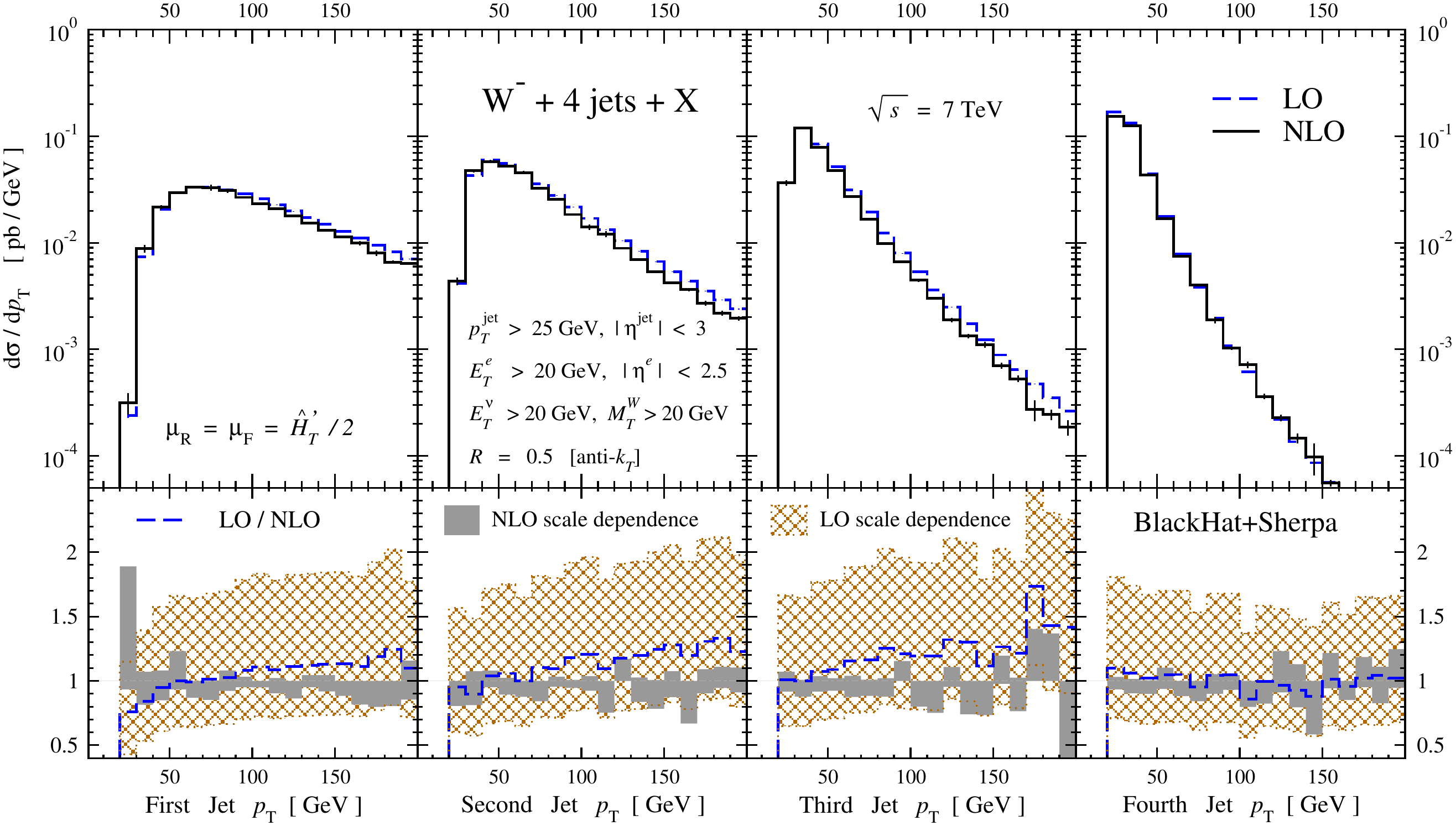}
  \caption{
	   Transverse momenta of the four leading jets in $pp\to W^-+4$ jets 
	   at the LHC calculated at \protect\NLO accuracy.
	   Figures taken from \cite{Berger:2010zx}.
	   \label{Fig:NLO_Wnjets_jetpt}
	  }
\end{figure}

Fig.\ \ref{Fig:NLO_Wnjets_jetpt} shows a calculation of the transverse 
momenta of four leading jets in $pp\to W+4$ jets production at the LHC 
using \BlackHat{}+\Sherpa \cite{Berger:2010zx,Gleisberg:2008ta} and a clear 
reduction of the theoretical uncertainty can be seen. However, possibly 
large logarithms due to scale hierarchies are not taken into account by 
such a calculation and also small scale dynamics are absent.

\subsection*{\protect\NLOPS}

To enhance next-to-leading order calculations with the resummation of 
large logarithms associated with the production of the softest jet 
it can be matched to a parton shower, either via the \MCatNLO 
\cite{Frixione:2002ik,Frixione:2003ei,Hoeche:2011fd} 
or the \POWHEG \cite{Nason:2004rx,Frixione:2007vw,Hamilton:2008pd,
  Hoeche:2010pf} technique. Such a calculation has the added benefit 
that it can take advantage of the parton shower's infrared continuation 
with the non-perturbative dynamics of hadronisation models with 
subsequent hadron decays, and multiple parton interactions. Thus, such 
calculation can directly be compared to experimental data.

\begin{figure}[t!]
  \begin{minipage}{0.48\columnwidth}
    \lineskip-1.8pt
    \includegraphics[width=\textwidth]{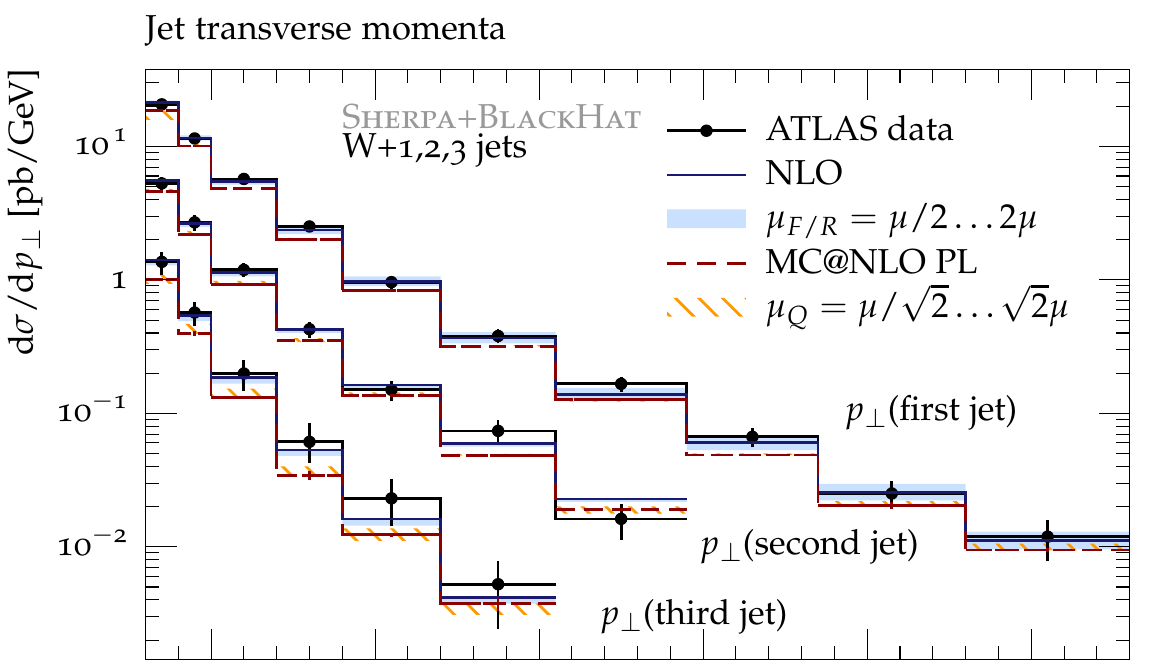}\\
    \includegraphics[width=\textwidth]{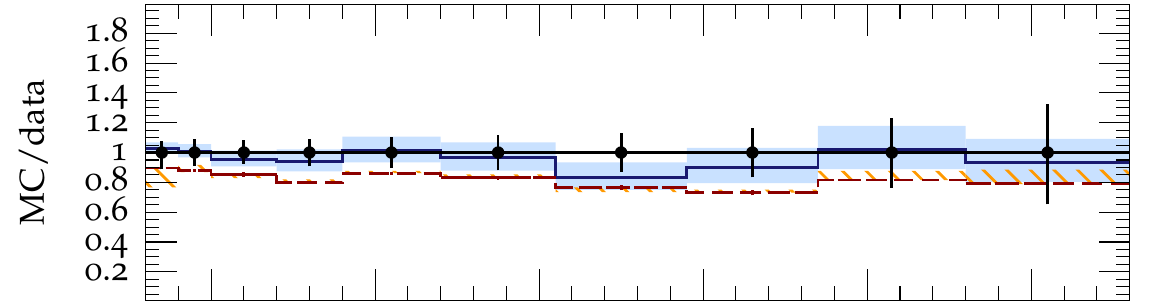}\\
    \includegraphics[width=\textwidth]{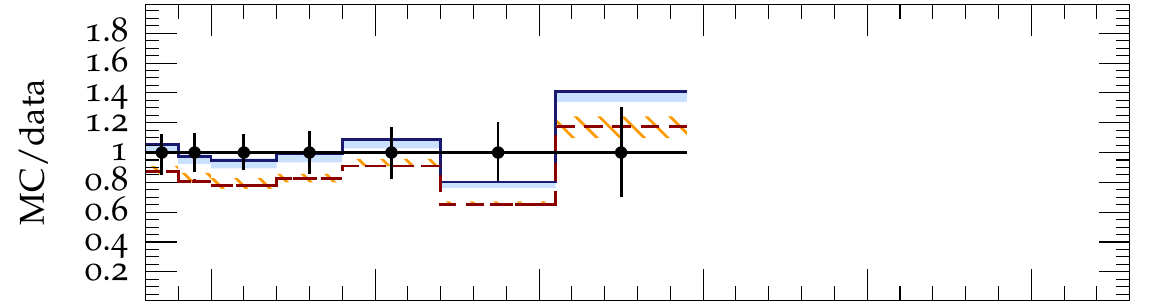}\\
    \includegraphics[width=\textwidth]{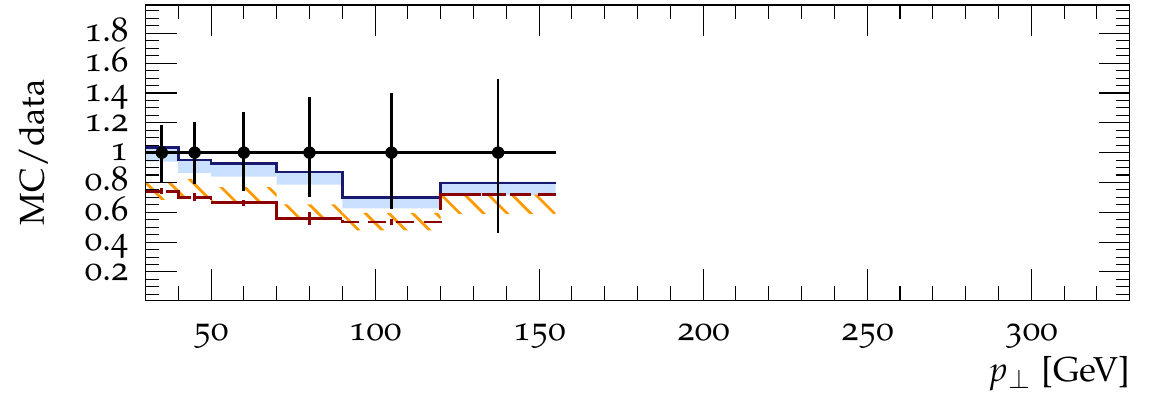}
  \end{minipage}\hfill
  \begin{minipage}{0.48\columnwidth}
    \includegraphics[width=\textwidth]{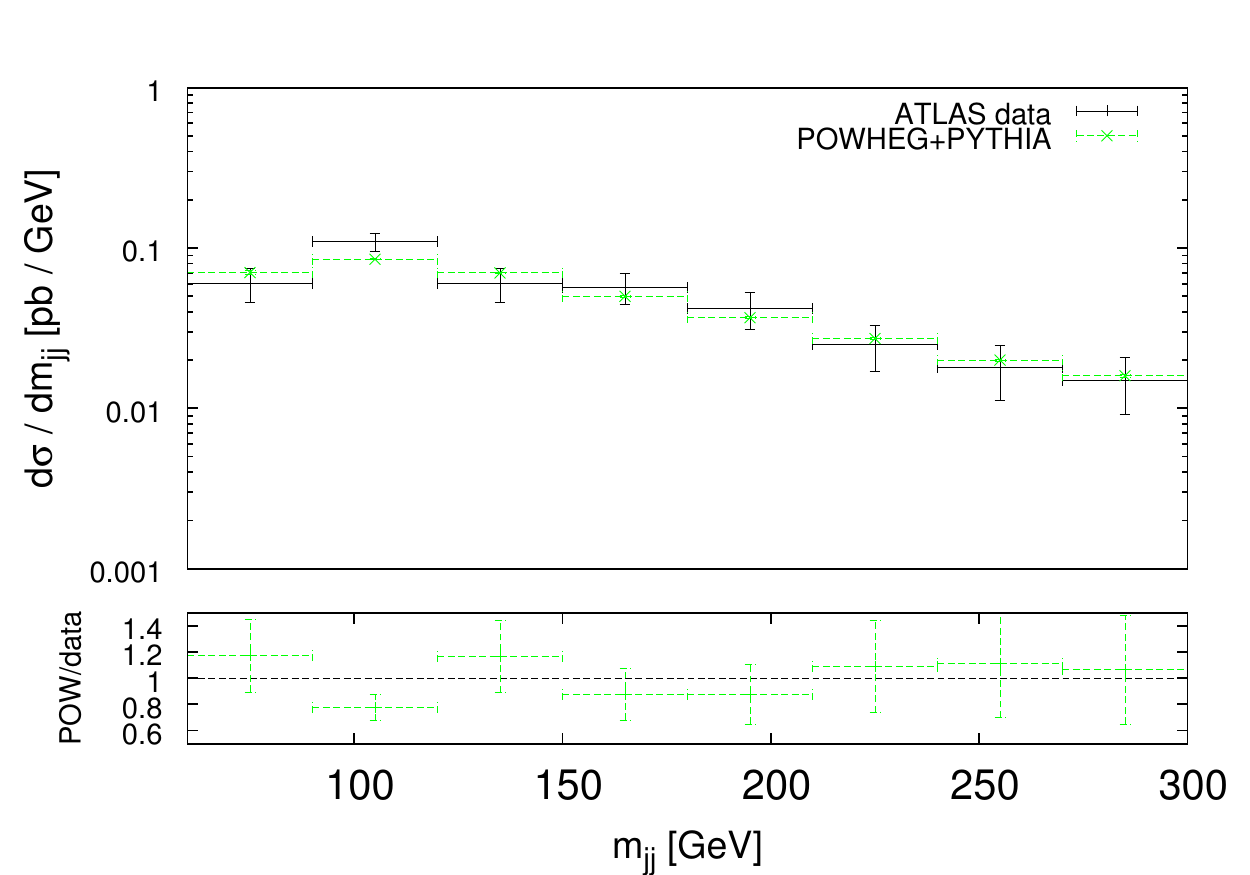}\vspace*{-40mm}\\
    \hspace*{-11mm}
    \includegraphics[width=1.2\textwidth]{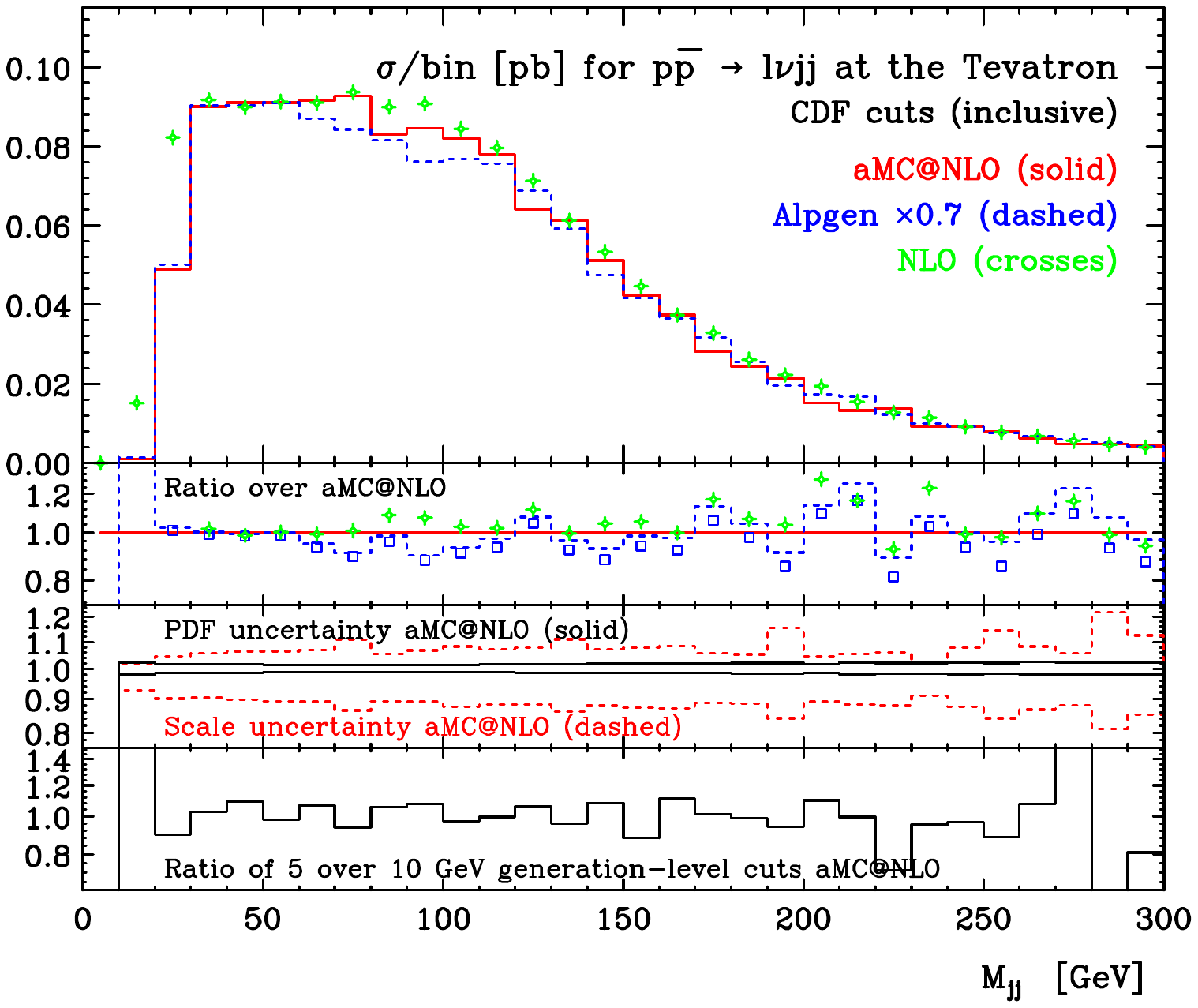}\vspace*{-24mm}
  \end{minipage}
  \caption{{\bf Left:} Transverse momenta of the $n$th jet in \MCatNLO 
	   simulations of $pp\to W+n$ jets at the LHC compared to \ATLAS 
	   data \cite{Aad:2012en}.
	   {\bf Right top:} Invariant mass of the leading jet pair in a 
	   \POWHEG simulation of $pp\to Z+2$ jets at the LHC compared to 
	   \ATLAS data \cite{Aad:2011qv}.
	   {\bf Right bottom:} Invariant mass of the leading jet pair 
	   in a \aMCatNLO simulation of $p\bar p\to W+2$ jets at the Tevatron.
	   Figures taken from \cite{Hoeche:2012ft,Re:2012zi,Frederix:2011ig}
	   \label{Fig:NLOPS_Wnjets_various}
	  }
\end{figure}

Fig.\ \ref{Fig:NLOPS_Wnjets_various} shows the results of implementations 
of these methods by various groups, reaching up to multiplicities of 
$pp\to V+3$ jets described at \NLO. For all, good agreement with data 
is found and the respective uncertainties are reduced compared to \LO 
calculations.

\subsection*{\protect\MEPSatNLO}

\NLOPS calculations, like fixed-order \NLO calculations, are lacking 
any resummation, of scale hierarchies of jet emissions that are 
described at \NLO accuracy, e.g.\ of the emission scales of the four 
leading jets in $pp\to V+4$ jets. Such a resummation of scales with 
respect to the inclusive sample, however, are present in \MEPS 
methods. Therefore, as a first step, \NLOPS calculations were combined 
with \MEPS merging methods to the called \MENLOPS merging method 
\cite{Hamilton:2010wh,Hoeche:2010kg}. Therein the most inclusive 
process is described by a \NLOPS matched calculation while leading 
order matrix elements are merged on top of it.

\MEPSatNLO merging \cite{Lavesson:2008ah,Hoeche:2012yf,Gehrmann:2012yg} 
now aims at merging \NLOPS matched calculations of successive jet 
multiplicities into such an inclusive sample. Therein, not only are the 
respective jet multiplicities described at next-to-leading order accuracy, 
but also the overall resummation of the parton shower is undisturbed. 
Thus, both large and small scale dynamics are accurately described.

\begin{figure}[t]
  \includegraphics[width=0.48\textwidth]{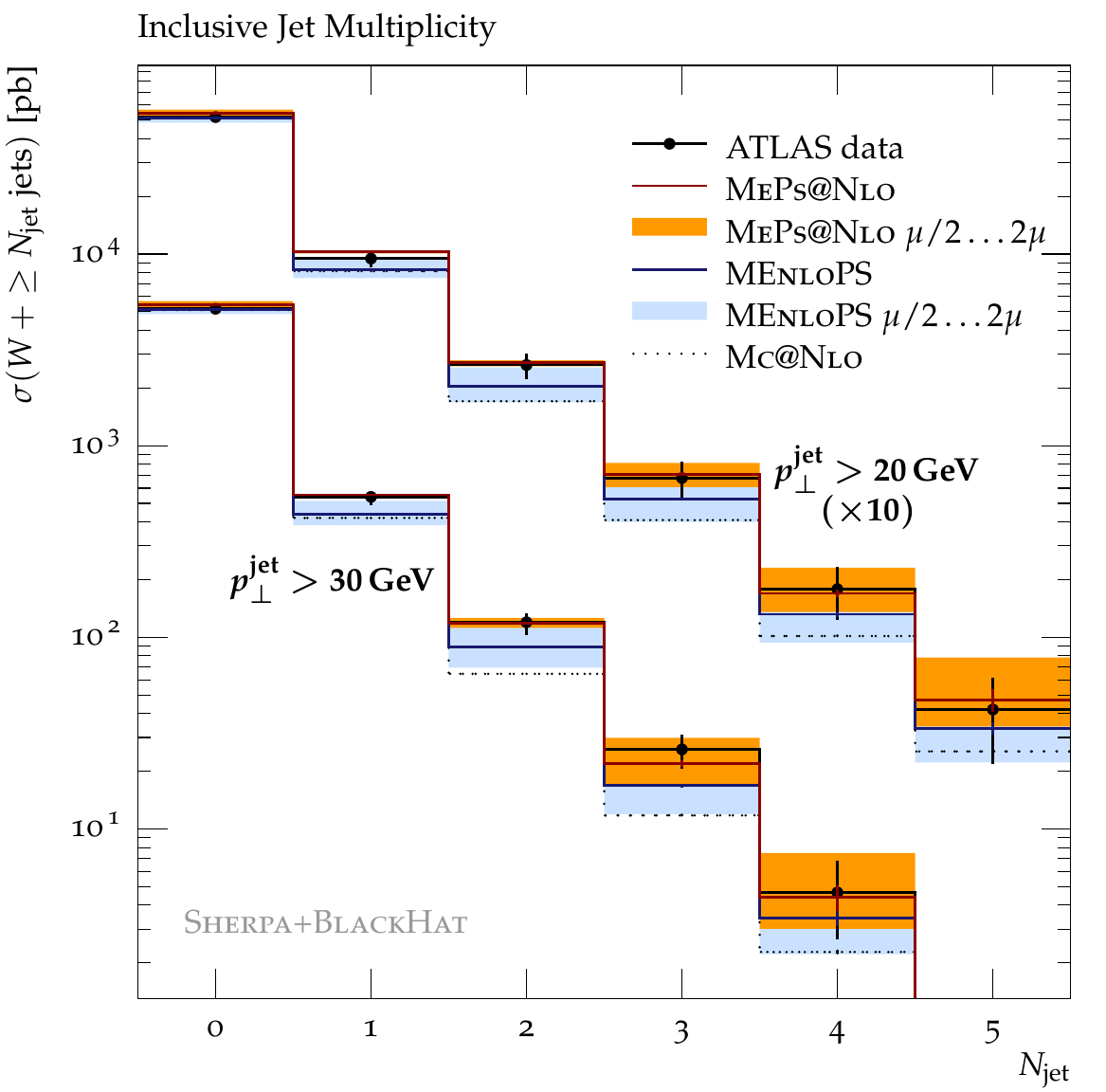}\hfill
  \includegraphics[width=0.48\textwidth]{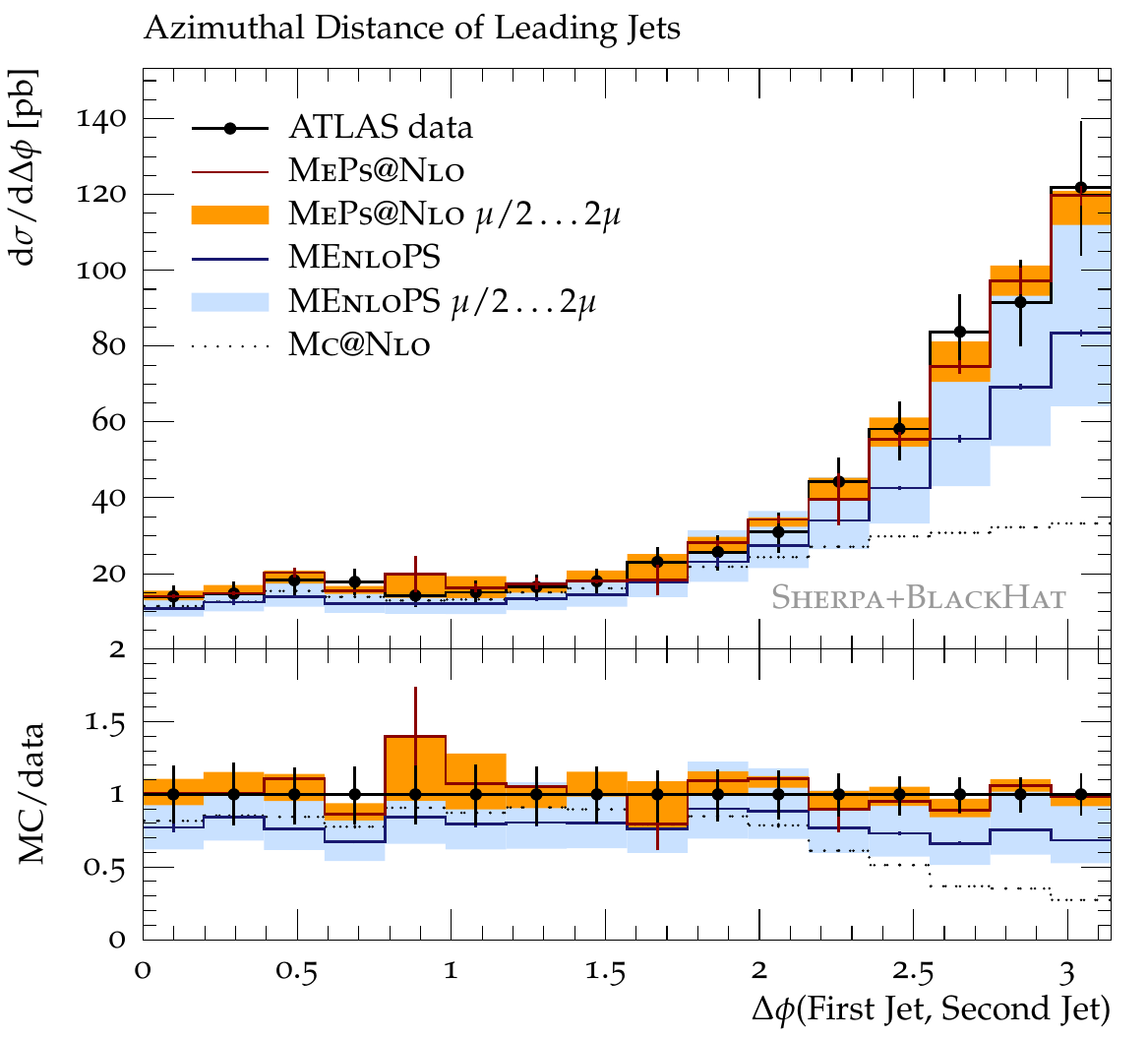}
  \caption{
	   $W+\ge n$-jet cross sections {\bf (left)} and azimuthal 
	   decorrellation of the two leading jets {\bf (right)} in 
	   $pp\to W+$ jets production at the LHC compared to \ATLAS 
	   data \cite{Aad:2012en}.
	   Figures taken from \cite{Hoeche:2012yf}.
	   \label{Fig:MEPSatNLO_Wjets_jetxs_dphi}
	  }
\end{figure}

\begin{figure}[t]
  \begin{minipage}{0.48\columnwidth}
    \lineskip-1.8pt
    \includegraphics[width=\textwidth]{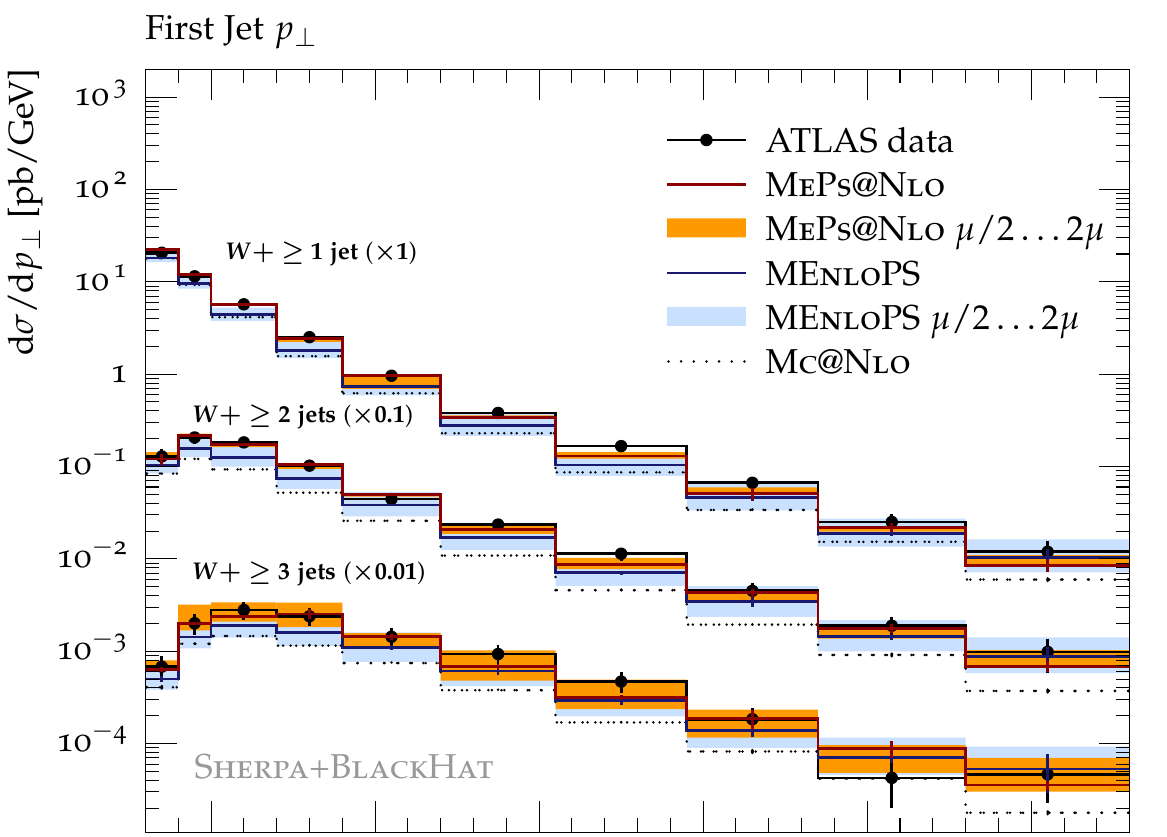}\\
    \includegraphics[width=\textwidth]{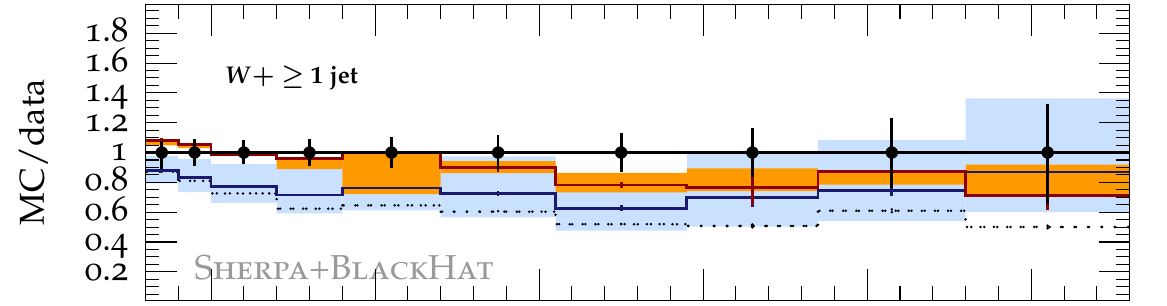}\\
    \includegraphics[width=\textwidth]{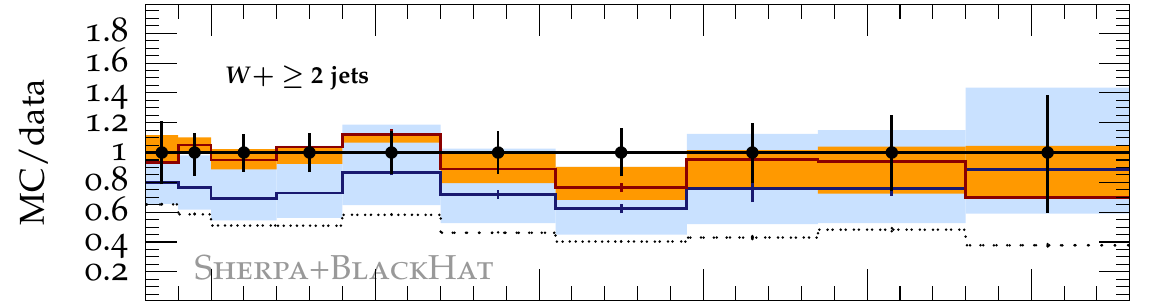}\\
    \includegraphics[width=\textwidth]{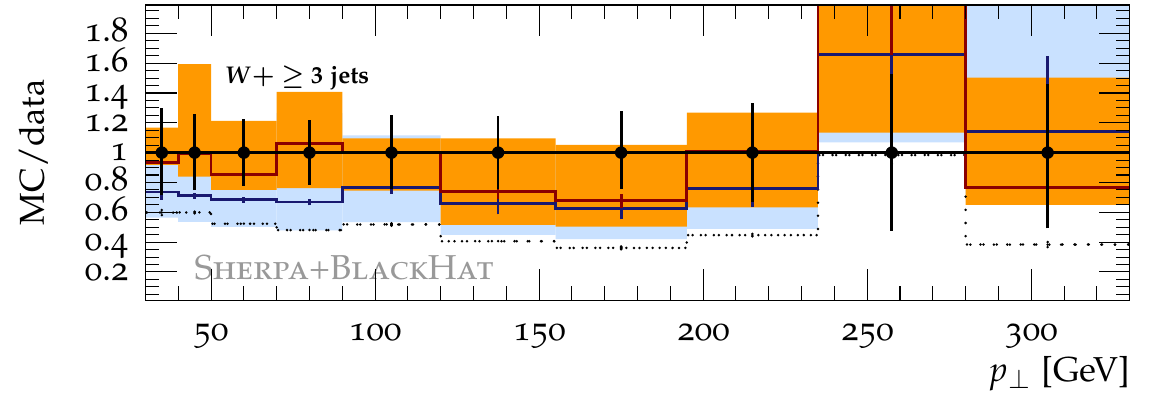}
  \end{minipage}\hfill
  \begin{minipage}{0.48\columnwidth}
    \lineskip-1.8pt
    \includegraphics[width=\textwidth]{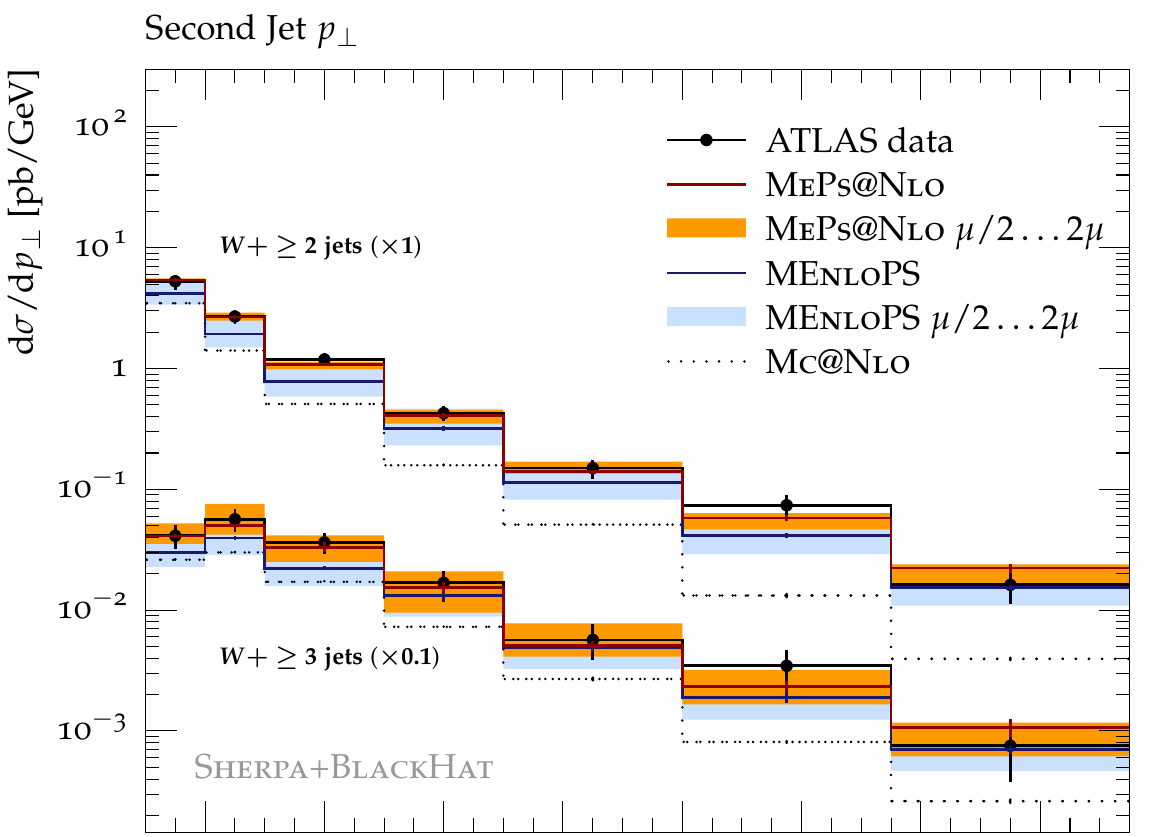}\\
    \includegraphics[width=\textwidth]{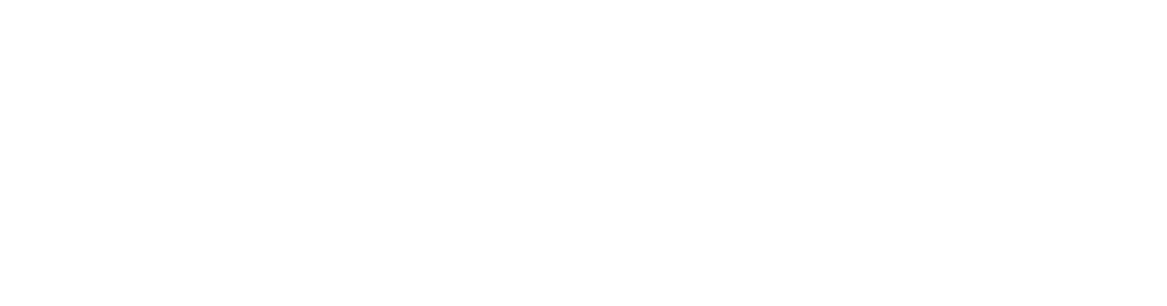}\\
    \includegraphics[width=\textwidth]{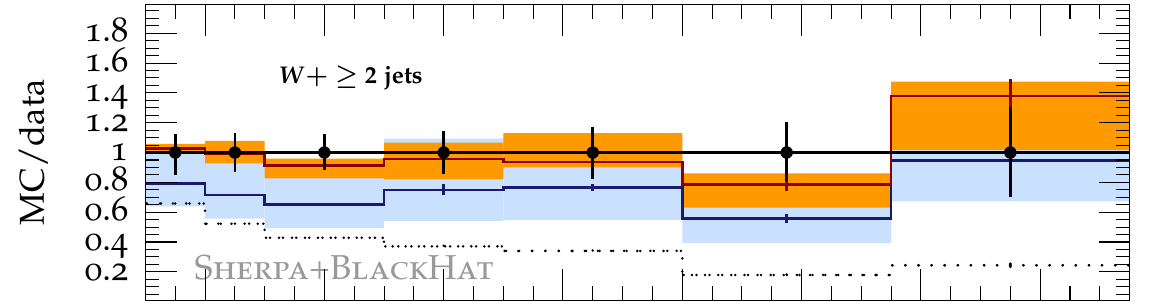}\\
    \includegraphics[width=\textwidth]{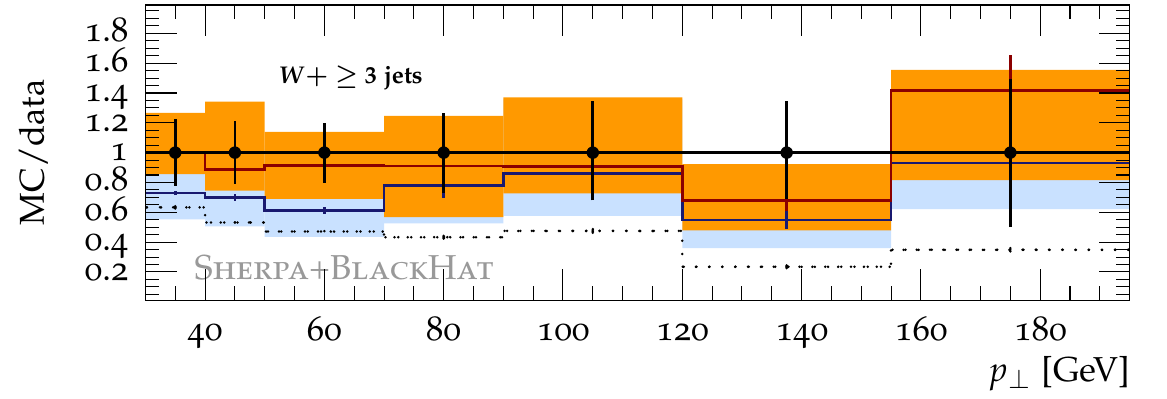}
  \end{minipage}
  \caption{
	   Transverse momentum of the leading and subleading jet in events 
	   with at least 1, 2 or 3 jets in $pp\to W+$ jets production at 
	   the LHC compare to \ATLAS data \cite{Aad:2012en}.
	   Figures taken from \cite{Hoeche:2012yf}.
	   \label{Fig:MEPSatNLO_Wjets_jetpt}
	  }
\end{figure}

Fig.\ \ref{Fig:MEPSatNLO_Wjets_jetxs_dphi} displays the inclusive $n$-jet 
cross sections and the azimuthal decorrelation of the two leading jets, 
probing both relative production rates and interjet dynamics. Comparing 
the uncertainties of the \MEPSatNLO merging method (merging $pp\to W+0,1,2$ 
jets at \NLO and $pp\to W+3,4$ jets at \LO) to those of the \MENLOPS method 
(merging $pp\to W+0$ jets at \NLO and $pp\to W+1,2,3,4$ jets at \LO) one 
clearly seas the added precision of including more \NLO matrix elements.
The predictions are compared to \ATLAS data \cite{Aad:2012en} and also 
the more accurate method is favoured.
Fig.\ \ref{Fig:MEPSatNLO_Wjets_jetpt} presents the transverse momenta of 
the two leading jets in events with at least one, two or three jets. 
A similar reduction of the uncertainties is found.

Finally, with these methods at hand, a full assertion of all perturbative 
and non-perturbative uncertainties in particle level Monte-Carlo 
predictions following \cite{Hoeche:2012fm} can be done.

\section*{References}
\bibliographystyle{iopart-num}  
\bibliography{journal}

\providecommand{\newblock}{}
\begin{thebibliography}{10}
\expandafter\ifx\csname url\endcsname\relax
  \def\url#1{{\tt #1}}\fi
\expandafter\ifx\csname urlprefix\endcsname\relax\def\urlprefix{URL }\fi
\providecommand{\eprint}[2][]{\url{#2}}

\bibitem{Catani:2001cc}
Catani S, Krauss F, Kuhn R and Webber B~R 2001 {\em JHEP\/} {\bf 11} 063
  (\textit{Preprint} \eprint{hep-ph/0109231})
  \urlprefix\url{http://www.slac.stanford.edu/spires/find/hep/www?eprint=hep-p%
h/0109231}

\bibitem{Mangano:2001xp}
Mangano M~L, Moretti M and Pittau R 2002 {\em Nucl. Phys.\/} {\bf B632}
  343--362 (\textit{Preprint} \eprint{hep-ph/0108069})
  \urlprefix\url{http://www.slac.stanford.edu/spires/find/hep/www?eprint=hep-p%
h/0108069}

\bibitem{Lonnblad:2001iq}
L{\"o}nnblad L 2002 {\em JHEP\/} {\bf 05} 046 (\textit{Preprint}
  \eprint{hep-ph/0112284})
  \urlprefix\url{http://www.slac.stanford.edu/spires/find/hep/www?eprint=hep-p%
h/0112284}

\bibitem{Krauss:2002up}
Krauss F 2002 {\em JHEP\/} {\bf 0208} 015 (\textit{Preprint}
  \eprint{hep-ph/0205283})
  \urlprefix\url{http://www.slac.stanford.edu/spires/find/hep/www?eprint=hep-p%
h/0205283}

\bibitem{Mangano:2006rw}
Mangano M~L, Moretti M, Piccinini F and Treccani M 2007 {\em JHEP\/} {\bf 01}
  013 (\textit{Preprint} \eprint{hep-ph/0611129})
  \urlprefix\url{http://www.slac.stanford.edu/spires/find/hep/www?eprint=hep-p%
h/0611129}

\bibitem{Alwall:2007fs}
Alwall J {\em et~al.\/} 2008 {\em Eur. Phys. J.\/} {\bf C53} 473--500
  (\textit{Preprint} \eprint{0706.2569})
  \urlprefix\url{http://www.slac.stanford.edu/spires/find/hep/www?eprint=arXiv%
:0706.2569}

\bibitem{Hoeche:2009rj}
H{\"o}che S, Krauss F, Schumann S and Siegert F 2009 {\em JHEP\/} {\bf 05} 053
  (\textit{Preprint} \eprint{0903.1219})
  \urlprefix\url{http://www.slac.stanford.edu/spires/find/hep/www?eprint=arXiv%
:0903.1219}

\bibitem{Hamilton:2009ne}
Hamilton K, Richardson P and Tully J 2009 {\em JHEP\/} {\bf 11} 038
  (\textit{Preprint} \eprint{0905.3072})
  \urlprefix\url{http://www.slac.stanford.edu/spires/find/hep/www?eprint=arXiv%
:0905.3072}

\bibitem{Lonnblad:2011xx}
L{\"o}nnblad L and Prestel S 2012 {\em JHEP\/} {\bf 03} 019 (\textit{Preprint}
  \eprint{1109.4829})

\bibitem{Aad:2010ab}
Aad G {\em et~al.\/} (ATLAS Collaboration) 2011 {\em Phys.Lett.\/} {\bf B698}
  325--345 (\textit{Preprint} \eprint{1012.5382})

\bibitem{Aad:2012en}
Aad G {\em et~al.\/} (ATLAS Collaboration) 2012 {\em Phys.Rev.\/} {\bf D85}
  092002 (\textit{Preprint} \eprint{1201.1276})
  \urlprefix\url{http://inspirehep.net/record/1083318}

\bibitem{Aad:2011qv}
Aad G {\em et~al.\/} (ATLAS) 2012 {\em Phys.Rev.\/} {\bf D85} 032009
  (\textit{Preprint} \eprint{1111.2690})
  \urlprefix\url{http://inspirehep.net/record/945498}

\bibitem{Chatrchyan:2011ne}
Chatrchyan S {\em et~al.\/} (CMS) 2012 {\em JHEP\/} {\bf 1201} 010
  (\textit{Preprint} \eprint{1110.3226})
  \urlprefix\url{http://inspirehep.net/record/940012}

\bibitem{:2013is}
Chatrchyan S {\em et~al.\/} (CMS Collaboration) 2013  (\textit{Preprint}
  \eprint{1301.1646})

\bibitem{Campbell:2002tg}
Campbell J~M and Ellis R 2002 {\em Phys. Rev.\/} {\bf D65} 113007
  (\textit{Preprint} \eprint{hep-ph/0202176})

\bibitem{Ellis:2009zw}
Ellis R, Melnikov K and Zanderighi G 2009 {\em JHEP\/} {\bf 0904} 077
  (\textit{Preprint} \eprint{0901.4101})

\bibitem{Berger:2009zg}
Berger C~F, Bern Z, Dixon L~J, Febres-Cordero F, Forde D, Gleisberg T, Ita H,
  Kosower D~A and Ma{\^i}tre D 2009 {\em Phys. Rev. Lett.\/} {\bf 102} 222001
  (\textit{Preprint} \eprint{0902.2760})
  \urlprefix\url{http://www.slac.stanford.edu/spires/find/hep/www?eprint=arXiv%
:0902.2760}

\bibitem{Berger:2010zx}
Berger C~F, Bern Z, Dixon L~J, Febres-Cordero F, Forde D, Gleisberg T, Ita H,
  Kosower D~A and Ma{\^i}tre D 2011 {\em Phys. Rev. Lett.\/} {\bf 106} 092001
  (\textit{Preprint} \eprint{1009.2338})
  \urlprefix\url{http://inspirebeta.net/record/867513}

\bibitem{Berger:2010vm}
Berger C~F, Bern Z, Dixon L~J, Febres-Cordero F, Forde D, Gleisberg T, Ita H,
  Kosower D~A and Ma{\^i}tre D 2010 {\em Phys. Rev.\/} {\bf D82} 074002
  (\textit{Preprint} \eprint{1004.1659})
  \urlprefix\url{http://www-spires.dur.ac.uk/spires/find/hep/www?eprint=arXiv:%
1004.1659}

\bibitem{Ita:2011wn}
Ita H, Bern Z, Dixon L~J, Febres-Cordero F, Kosower D~A and Ma{\^i}tre D 2012
  {\em Phys.Rev.\/} {\bf D85} 031501 (\textit{Preprint} \eprint{1108.2229})
  \urlprefix\url{http://inspirehep.net/record/922868}

\bibitem{Gleisberg:2008ta}
Gleisberg T, H{\"o}che S, Krauss F, Sch\"{o}nherr M, Schumann S, Siegert F and
  Winter J 2009 {\em JHEP\/} {\bf 02} 007 (\textit{Preprint}
  \eprint{0811.4622}) \urlprefix\url{http://inspirebeta.net/record/803708}

\bibitem{Frixione:2002ik}
Frixione S and Webber B~R 2002 {\em JHEP\/} {\bf 06} 029 (\textit{Preprint}
  \eprint{hep-ph/0204244})
  \urlprefix\url{http://www.slac.stanford.edu/spires/find/hep/www?eprint=hep-p%
h/0204244}

\bibitem{Frixione:2003ei}
Frixione S, Nason P and Webber B~R 2003 {\em JHEP\/} {\bf 08} 007
  (\textit{Preprint} \eprint{hep-ph/0305252})
  \urlprefix\url{http://www.slac.stanford.edu/spires/find/hep/www?eprint=hep-p%
h/0305252}

\bibitem{Hoeche:2011fd}
H{\"o}che S, Krauss F, Sch{\"o}nherr M and Siegert F 2012 {\em JHEP\/} {\bf 09}
  049 (\textit{Preprint} \eprint{1111.1220})
  \urlprefix\url{http://inspirehep.net/record/944643}

\bibitem{Nason:2004rx}
Nason P 2004 {\em JHEP\/} {\bf 11} 040 (\textit{Preprint}
  \eprint{hep-ph/0409146}) \urlprefix\url{http://inspirebeta.net/record/659055}

\bibitem{Frixione:2007vw}
Frixione S, Nason P and Oleari C 2007 {\em JHEP\/} {\bf 11} 070
  (\textit{Preprint} \eprint{0709.2092})
  \urlprefix\url{http://www.slac.stanford.edu/spires/find/hep/www?eprint=arXiv%
:0709.2092}

\bibitem{Hamilton:2008pd}
Hamilton K, Richardson P and Tully J 2008 {\em JHEP\/} {\bf 10} 015
  (\textit{Preprint} \eprint{0806.0290})
  \urlprefix\url{http://www.slac.stanford.edu/spires/find/hep/www?eprint=arXiv%
:0806.0290}

\bibitem{Hoeche:2010pf}
H{\"o}che S, Krauss F, Sch{\"o}nherr M and Siegert F 2011 {\em JHEP\/} {\bf 04}
  024 (\textit{Preprint} \eprint{1008.5399})
  \urlprefix\url{http://inspirebeta.net/record/866705}

\bibitem{Hoeche:2012ft}
H{\"o}che S, Krauss F, Sch{\"o}nherr M and Siegert F 2013 {\em
  Phys.Rev.Lett.\/} {\bf 110} 052001 (\textit{Preprint} \eprint{1201.5882})
  \urlprefix\url{http://inspirehep.net/record/1086175}

\bibitem{Re:2012zi}
Re E 2012 {\em JHEP\/} {\bf 1210} 031 (\textit{Preprint} \eprint{1204.5433})
  \urlprefix\url{http://inspirehep.net/record/1112186}

\bibitem{Frederix:2011ig}
Frederix R, Frixione S, Hirschi V, Maltoni F, Pittau R {\em et~al.\/} 2012 {\em
  JHEP\/} {\bf 1202} 048 (\textit{Preprint} \eprint{1110.5502})

\bibitem{Hamilton:2010wh}
Hamilton K and Nason P 2010 {\em JHEP\/} {\bf 06} 039 (\textit{Preprint}
  \eprint{1004.1764})
  \urlprefix\url{http://www.slac.stanford.edu/spires/find/hep/www?eprint=arXiv%
:1004.1764}

\bibitem{Hoeche:2010kg}
H{\"o}che S, Krauss F, Sch{\"o}nherr M and Siegert F 2011 {\em JHEP\/} {\bf 08}
  123 (\textit{Preprint} \eprint{1009.1127})
  \urlprefix\url{http://www.slac.stanford.edu/spires/find/hep/www?eprint=arXiv%
:1009.1127}

\bibitem{Lavesson:2008ah}
Lavesson N and L{\"o}nnblad L 2008 {\em JHEP\/} {\bf 12} 070 (\textit{Preprint}
  \eprint{0811.2912})

\bibitem{Hoeche:2012yf}
H{\"o}che S, Krauss F, Sch{\"o}nherr M and Siegert F 2012  (\textit{Preprint}
  \eprint{1207.5030}) \urlprefix\url{http://inspirehep.net/record/1123387}

\bibitem{Gehrmann:2012yg}
Gehrmann T, H{\"o}che S, Krauss F, Sch{\"o}nherr M and Siegert F 2013 {\em
  JHEP\/} {\bf 1301} 144 (\textit{Preprint} \eprint{1207.5031})

\bibitem{Hoeche:2012fm}
H{\"o}che S and Sch{\"o}nherr M 2012 {\em Phys.Rev.\/} {\bf D86} 094042
  (\textit{Preprint} \eprint{1208.2815})
  \urlprefix\url{http://inspirehep.net/record/1127523}

\end{thebibliography}
\end{document}